\documentclass[prl,showpacs,superscriptaddress]{revtex4}
\usepackage[ansinew]{inputenc}
\usepackage{graphicx}
\usepackage{hyperref}
\usepackage{amsmath}
\usepackage{amsthm}
\usepackage{bm}
\usepackage{layout}
\usepackage{float}
\usepackage{amsfonts}
\usepackage{amssymb}%
\usepackage[ansinew]{inputenc}
\usepackage{graphicx}
\usepackage{amsmath}
\usepackage{amsthm}
\usepackage{bm}
\usepackage{layout}
\usepackage{float}
\usepackage{amsfonts}
\usepackage{amssymb}
\usepackage{bm}
\usepackage{graphicx}
\usepackage{float}
\usepackage{color}
\usepackage{appendix}

\begin{document}

\title{Comment on Adler's ``Does the Peres experiment using photons test for hyper-complex
(quaternionic) quantum theories?''}

\author{Lorenzo M. Procopio}
\affiliation{Faculty of Physics, University of Vienna,
Boltzmanngasse 5, A-1090 Vienna, Austria}
\author{Lee A. Rozema}
\affiliation{Faculty of Physics, University of Vienna,
Boltzmanngasse 5, A-1090 Vienna, Austria}
\author{Borivoje Daki\' c}
\affiliation{Faculty of Physics, University of Vienna,
Boltzmanngasse 5, A-1090 Vienna, Austria} \affiliation{Institute of
Quantum Optics and Quantum Information, Austrian Academy of
Sciences, Boltzmanngasse 3, A-1090 Vienna, Austria}
\author{Philip Walther}
\affiliation{Faculty of Physics, University of Vienna,
Boltzmanngasse 5, A-1090 Vienna, Austria}

\date{\today}

\begin{abstract}
In his recent article [arXiv:1604.04950], Adler questions the usefulness of the bound found in our experimental search for genuine effects of hyper-complex quantum mechanics [arXiv:1602.01624]. Our experiment was performed using a black-box (instrumentalist) approach to generalized probabilistic theories; therefore, it does not assume \textit{a priori} any particular underlying mechanism.
From that point of view our experimental results do indeed place meaningful bounds on possible effects of ``post-quantum theories'', including quaternionic quantum mechanics. In his article, Adler compares our experiment to non-relativistic and M\"oller formal scattering theory within quaternionic quantum mechanics. With a particular set of assumptions, he finds that quaternionic effects would likely not manifest themselves in general.
Although these assumptions are justified in the non-relativistic case, a proper calculation for relativistic particles is still missing.
Here, we provide a concrete relativistic example of Klein-Gordon scattering wherein the quaternionic effects persist.
We note that when the Klein-Gordon equation is formulated using a Hamiltonian formalism it displays a so-called ``indefinite metric'', a characteristic feature of relativistic quantum wave equations. In Adler's example this is directly forbidden by his assumptions, and therefore our present example is not in contradiction to his work. In complex quantum mechanics this problem of an indefinite metric is solved in second quantization. Unfortunately, there is no known algorithm for canonical field quantization in quaternionic quantum mechanics.
\end{abstract}
\maketitle

In the instrumentalist approach to a probabilistic theory \cite{Hardy2001,barrett2007information}, a physical system and physical states are described by any set of mathematical objects that correctly predict experimental results. Such a framework considers experimental situations as a manipulation of primitive laboratory devices such as preparations, transformations, and measurements. No assumption is \textit{a priori} made about any potential underlying mechanism that may be used to characterize the more elementary processes taking place within each laboratory device. From that perspective both quantum theory and its hyper-complex generalizations~\cite{Jordan,adler1995quaternionic} are just particular theories in a vast sea of so-called generalized probabilistic theories \cite{Hardy2001,barrett2007information}. Whether nature prefers to use one theory or another is then an experimental issue---unless one postulates a set of axioms to single out one particular theory.

Our recent experiment \cite{Procopio2016} should be understood purely within such an operational framework.
Namely, in our experiment we use physically different phase shifters as transformation devices and, in our analysis, we treat them as black-boxes.
One may naively guess that they have no observable effect on a system sent through them.
However, in quantum theory such a phase shift can effect the outcomes of interference experiments, showing that these boxes perform non-trivial transformations.
One can ask further if the order of the boxes can produce any observable effect? Within complex quantum theory the answer is negative as these transformations are commutative, and hence different orderings cannot lead to an observable effect.
However, a detailed analysis shows that within the framework of generalized probabilistic theories such composite transformations are non-trivial~\cite{Muller}. Therefore, the commutativity of such boxes becomes an experimental issue. If such an effect does exist, one should be able to observe it unless the boxes are identical. The effect could be tiny; therefore, one may need a very precise measurement to detect it.
Viewed from this framework, our experiment provides such a precision measurement, and, to the best of our knowledge, it places the tightest explicit bound on such effects to date.

If one wishes to see how close our experimental test is to the deviation (if any) predicted by hyper-complex quantum theories one has to invoke a particular model to characterize interactions between the input system and the elementary constituents of the box.
In his model \cite{adler1995quaternionic,Adler2016} Adler shows that all the non-trivial quaternionic effects are exponentially suppressed for the non-relativistic case. Therefore, one should not expect to see any observable deviation in our experiment.
However, his calculation is carried out for the case where the ``free Hamiltonian'' (or ``asymptotic Hamiltonian'' in his terminology) $\tilde{H_0}$ has very specific properties. Namely, it is assumed that the free Hamiltonian, through its spectral decomposition and using the re-raying freedom of
quaternionic quantum theory~\cite{AdlerPrivate}, directly gives a preferred imaginary unit (i.e. all of the eigenvalues of $\tilde{H_0}$ share the same imaginary unit $i$, $\tilde{E}_n=i E_n$). This assumption turns out to be very important to derive the main claim. In the Appendix we provide a simple counterexample of the scattering of a ``Klein-Gordon wave'' a on a quaternionic potential. We show that there is no exponential damping in the transmitted wave. However, our example is not in contradiction with Adler's calculation, as the Klein-Gordon equation when formulated using a Hamiltonian formalism displays a so-called indefinite metric, a characteristic feature of relativistic quantum wave equations~\cite{FESHBACH}. In Adler's calculations this is directly forbidden by his assumptions. The problem of the indefinite metric is resolved within complex quantum mechanics by a second quantization. Unfortunately, the problem of quantizing relativistic fields in quaternionic quantum mechanics is still open~\cite{AdlerPrivate}.

Unlike the non-relativistic case for which Adler's model is justified by explicit calculations~\cite{AdlerNonRel1,adler1995quaternionic}  there is no complete relativistic treatment for massless particles such as single photons. One of reasons for this is that there is no known recipe for a full quantum-field approach to quanternionic quantum mechanics~\cite{adler1995quaternionic,AdlerPrivate}. Therefore the applicability of the particular form of quaternionic quantum mechanics shown in \cite{Adler2016} to our photonic experiment still remains in the domain of conjecture.

Nevertheless, the model of Adler suggests that perhaps one has to perform near field experiments to observe the quaternionic effects predicted by his formulation. We fully acknowledge this, and indeed it may well be a focus of future experiments.
In this light, our recent experiment should be understood not as the final word on quaternionic quantum mechanics but rather as a significant step towards an exhaustive experimental search for genuine post-quantum theories.

\emph{Acknowledgment.} We wish to thank \v Caslav Brukner and Stephen Adler for stimulating discussions.
\bibliography{quaternion_comment}

\section{Appendix: Scattering from a quaternionic rectangular-barrier}

In this section we calculate the solutions to the Klein-Gordon equation for a quaternionic rectangular potential.
We find that a non-decaying quaternionic phase remains, and this could lead to a deviation from complex quantum theory. The calculation of this Appendix, through the symplectic component matching equations of Eq. (11), is based
on notes sent to us by Stephen Adler~\cite{Adlernote}.

In the usual (complex) relativistic quantum mechanics, the interaction of the one-dimensional Klein-Gordon field $\phi(x,t)$ with the field $A^{\mu}(x,t)=(V(x),0)$ is described in a \emph{gauge invariant} way
\begin{equation}
\left[-\left(\frac{\partial}{\partial t} -i V(x) \right)^2 +\frac{\partial^2}{\partial x^2}\right]\phi(x,t)=m^2\phi(x,t),
\end{equation}
where we have set all the relevant constants to $c=\hbar=1$.
To extend this to quaternionic potentials one can assume the correspondence $i V(x)\mapsto \tilde{V}(x)$ where $\tilde{V}(x)$ is an imaginary quaternion field $\tilde{V}(x)^{*}=-\tilde{V}(x)$.

\begin{figure}
    \centering
    \includegraphics[scale=0.7]{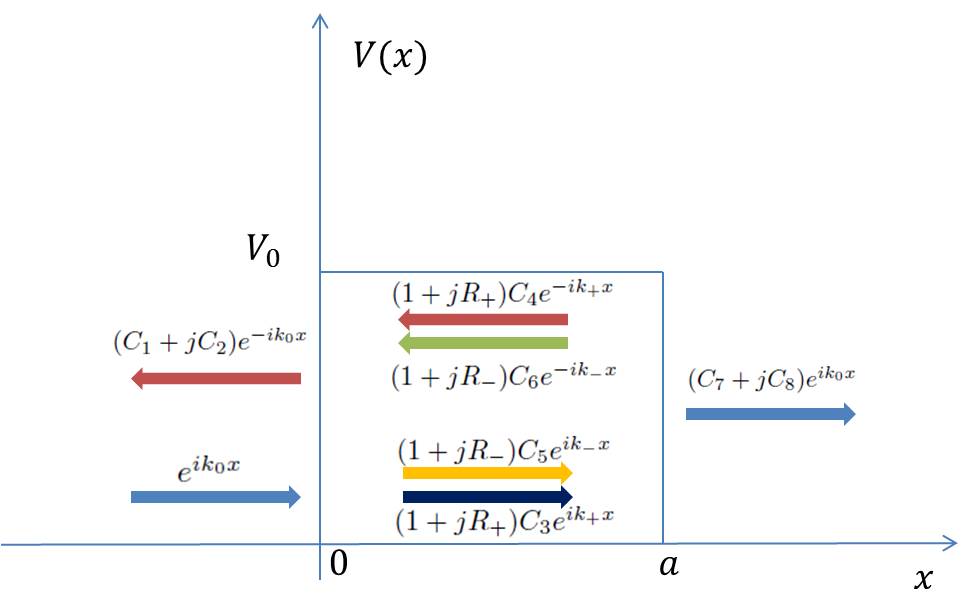}
    \caption{Simple sketch of the scattering scenario. An incident wave $e^{ik_0x}$ scatters off of a quaternionic rectangular potential resulting in one reflected wave $(C_1+jC_2)e^{-ik_0x}$  and one transmitted wave $(C_7+jC_8)e^{ik_0x}$.  In the text we show that transmitted/reflected amplitudes are quaternionic in general.}
    \label{barrier}
\end{figure}

Let consider a scattering process of a massless particle ($m=0$) on a rectangular-potential
\begin{equation}
\tilde{V}(x)=n \left\{
                                                               \begin{array}{ll}
                                                                 0, & \hbox{$x<0$;} \\
                                                                 V_0, & \hbox{$0\leq x\leq a$;} \\
                                                                 0, & \hbox{$x>0$.}
                                                               \end{array}
                                                             \right.
\end{equation}
with $n$ being an imaginary unit quaternion $n=n_1 i+n_2 j+n_3 k$; i.e. $n_1^2+n_2^2+n_3^2=1$ and $a, V_0$ are the barrier parameters. We start by solving the KG equation with the following ansatz: $\phi(x,t)=Ce^{i(kx-\omega_0t)}$, where $C$ is quaternionic in general. We get the following condition
\begin{equation}
\omega_0^2C-2\omega_0\tilde{V}(x)Ci-\tilde{V}(x)^2C-k^2C=0.
\end{equation}
Outside of barrier we have $\tilde{V}(x)=0$, hence there is a nontrivial solution iff $k=\pm\omega_0=\pm k_0$. For the region inside of barrier we have $\tilde{V}(x)=nV_0$. We write the coefficient $C$ in terms of its symplectic components $C=C_\alpha+jC_\beta$, where $C_{\alpha,\beta}$ are complex numbers. Having in mind that
\begin{eqnarray}
n(C_\alpha+jC_\beta)i&=&ni(C_\alpha-jC_\beta)\\\nonumber
&=&\left(-n_1+j(n_3+in_2)\right)(C_\alpha-jC_\beta)\\\nonumber
&=&-n_1C_\alpha+(n_3-in_2C_\beta)+j\left((n_3+in_2)C_\alpha-n_1C_\beta\right),
\end{eqnarray}
we get the following set of coupled equations
\begin{eqnarray}
(\omega_0^2+V_0^2+k^2-2\omega_0V_0n_1)C_\alpha-2\omega_0V_0(n_3-in_2)C_\beta&=&0\\\nonumber
-2\omega_0V_0(n_3+in_2)C_\alpha+(\omega_0^2+V_0^2+k^2+2\omega_0V_0n_1)C_\beta&=&0.
\end{eqnarray}
The system has a nontrivial solution if its determinant is set to zero, therefore we get
\begin{equation}
(\omega_0^2+V_0^2+k^2)^2-4\omega_0^2V_0^2(n_1^2+n_2^2+n_3^2)=(\omega_0^2+V_0^2+k^2)^2-4\omega_0^2V_0^2=0,
\end{equation}
or equivalently $k=\pm|\omega_0\pm V_0|=\pm k_{\pm}$. Consequently, we have
\begin{equation}
\frac{C_\beta}{C_\alpha}=-\frac{n_1\pm1}{n_3-in_2}=R_{\pm}.
\end{equation}
Now we can write the spatial part (complete solution reads $\phi(x,t)=\psi(x)e^{-i\omega_0t}$) of the wavefunction in the whole region
\begin{eqnarray}\label{xxx}
\psi(x)  &=& \begin{cases}e^{ik_0x} +(C_1+jC_2)e^{-ik_0x}, & x < 0,
   \\
(1+jR_+)C_3e^{ik_+x}+(1+jR_+)C_4e^{-ik_+x}+(1+jR_-)C_5e^{ik_-x}+(1+jR_-)C_6e^{-ik_-x}, & 0\leq x\leq a,\\
(C_7+jC_8)e^{ik_0x}, & x>a.
 \end{cases}
\end{eqnarray}
All the constants $C_i$ are complex numbers and we intentionally have set the amplitude of the incident beam $e^{ik_0x}$ to be 1. Both $\psi(x)$ and $\partial_x\psi(x)$  have to be continuous at $x=0$ and $x=a$. We have
\begin{eqnarray}\label{Dxxx}
\partial_x\psi(x)  &=& \begin{cases}e^{ik_0x}ik_0 -(C_1+jC_2)e^{-ik_0x}ik_0, & x < 0,
   \\
(1+jR_+)C_3e^{ik_+x}ik_+-(1+jR_+)C_4e^{-ik_+x}ik_++\\
+(1+jR_-)C_5e^{ik_-x}ik_--(1+jR_-)C_6e^{-ik_-x}ik_-, & 0\leq x\leq a,\\
(C_7+jC_8)e^{ik_0x}ik_0, & x>a.
 \end{cases}
\end{eqnarray}
The corresponding wave-matching equations are
\begin{eqnarray}
&&1+C_1+jC_2=(1+jR_+)C_3+(1+jR_+)C_4+(1+jR_-)C_5+(1+jR_-)C_6\\\nonumber
&&k_0 -(C_1+jC_2)k_0=(1+jR_+)C_3k_+-(1+jR_+)C_4k_++(1+jR_-)C_5k_--(1+jR_-)C_6k_-\\\nonumber
&&(1+jR_+)C_3e^{ik_+a}+(1+jR_+)C_4e^{-ik_+a}+(1+jR_-)C_5e^{ik_-a}+(1+jR_-)C_6e^{-ik_-a}=(C_7+jC_8)e^{ik_0a}\\\nonumber
&&(1+jR_+)C_3e^{ik_+a}k_+-(1+jR_+)C_4e^{-ik_+a}k_++(1+jR_-)C_5e^{ik_-a}k_--(1+jR_-)C_6e^{-ik_-a}k_-=(C_7+jC_8)e^{ik_0a}k_0,
\end{eqnarray}
or in terms of symplectic components we get the following set of equations
\begin{eqnarray}
&&1+C_1=C_3+C_4+C_5+C_6\\\nonumber
&&C_2=R_+C_3+R_+C_4+R_-C_5+R_-C_6\\\nonumber
&&k_0 -C_1k_0=C_3k_+-C_4k_++C_5k_--C_6k_-\\\nonumber
&&-C_2k_0=R_+C_3k_+-R_+C_4k_++R_-C_5k_--R_-C_6k_-\\\nonumber
&&C_3e^{ik_+a}+C_4e^{-ik_+a}+C_5e^{ik_-a}+C_6e^{-ik_-a}=C_7e^{ik_0a}\\\nonumber
&&R_+C_3e^{ik_+a}+R_+C_4e^{-ik_+a}+R_-C_5e^{ik_-a}+R_-C_6e^{-ik_-a}=C_8e^{ik_0a}\\\nonumber
&&C_3e^{ik_+a}k_+-C_4e^{-ik_+a}k_++C_5e^{ik_-a}k_--C_6e^{-ik_-a}k_-=C_7e^{ik_0a}k_0\\\nonumber
&&R_+C_3e^{ik_+a}k_+-R_+C_4e^{-ik_+a}k_++R_-C_5e^{ik_-a}k_--R_-C_6e^{-ik_-a}k_-=C_8e^{ik_0a}k_0.
\end{eqnarray}
The system above can be written in matrix form $M\vec{C}=\vec{C_0}$, where
\begin{equation}
M=\left(
\begin{array}{cccccccc}
 1 & 0 & -1 & -1 & -1 & -1 & 0 & 0 \\
 0 & 1 & -R_+ & -R_+ & -R_- & -R_- & 0 & 0 \\
 -k_0 & 0 & -k_+ & k_+ & -k_- & k_- & 0 & 0 \\
 0 & -k_0 & -k_+ R_+ & k_+ R_+ & -k_- R_- & k_- R_- & 0 & 0 \\
 0 & 0 & e^{i a k_+} & e^{-i a k_+} & e^{i a k_-} & e^{-i a k_-} & -e^{i a k_0} & 0 \\
 0 & 0 & e^{i a k_+} R_+ & e^{-i a k_+} R_+ & e^{i a k_-} R_- & e^{-i a k_-} R_- & 0 & -e^{i a k_0} \\
 0 & 0 & e^{i a k_+} k_+ & -e^{-i a k_+} k_+ & e^{i a k_-} k_- & -e^{-i a k_-} k_- & -e^{i a k_0} k_0 & 0 \\
 0 & 0 & e^{i a k_+} k_+ R_+ & -e^{-i a k_+} k_+ R_+ & e^{i a k_-} k_- R_- & -e^{-i a k_-} k_- R_- & 0 & -e^{i a k_0} k_0 \\
\end{array}
\right)
\end{equation}
and $\vec{C_0}=-(1,0,k_0,0,0,0,0,0)^{T}$.

Using Mathematica software we obtained the following set of solutions
\begin{eqnarray}
C_1&=&-\frac{\left(k_-^2-k_0^2\right) \sin \left(a k_-\right)}{k_-^2 \sin \left(a k_-\right)+k_0^2 \sin \left(a k_-\right)+2 i k_0 k_- \cos \left(a
   k_-\right)}\frac{R_+}{R_+-R_-}+\\\nonumber
&+&\frac{\left(k_+^2-k_0^2\right) \sin \left(a k_+\right)}{k_+^2 \sin \left(a k_+\right)+k_0^2 \sin \left(a k_+\right)+2 i k_0 k_+ \cos \left(a
   k_+\right)}\frac{R_-}{R_+-R_-},\\
C_2&=&\left(\frac{\left(k_+^2-k_0^2\right) \sin \left(a k_+\right)}{\left(k_+^2+k_0^2\right) \sin \left(a k_+\right)+2 i k_+ k_0 \cos \left(a
   k_+\right)}+\frac{\left(k_0^2-k_-^2\right) \sin \left(a k_-\right)}{\left(k_-^2+k_0^2\right) \sin \left(a k_-\right)+2 i k_- k_0 \cos
   \left(a k_-\right)}\right)\frac{R_+R_-}{R_+-R_-},\\
C_3&=&\frac{2 k_0 \left(k_++k_0\right)}{-\left(k_++k_0\right){}^2+\left(k_+-k_0\right){}^2 e^{2 i a k_+}}\frac{R_-}{R_+-R_-},\\
C_4&=&\frac{2 \left(k_+-k_0\right) k_0 e^{2 i a k_+}}{-\left(k_++k_0\right){}^2+\left(k_+-k_0\right){}^2 e^{2 i a k_+}}\frac{R_-}{R_+-R_-},\\
C_5&=&-\frac{2 k_0 \left(k_-+k_0\right)}{-\left(k_-+k_0\right){}^2+\left(k_--k_0\right){}^2 e^{2 i a k_-}}\frac{R_+}{R_+-R_-},\\
C_6&=&-\frac{2 \left(k_--k_0\right) k_0 e^{2 i a k_-}}{-\left(k_-+k_0\right){}^2+\left(k_--k_0\right){}^2 e^{2 i a k_-}}\frac{R_+}{R_+-R_-},\\\nonumber
C_7&=&-\frac{2 i k_+ k_0 e^{-i a k_0}}{k_+^2 \sin \left(a k_+\right)+k_0^2 \sin \left(a k_+\right)+2 i k_0 k_+ \cos \left(a k_+\right)}\frac{R_-}{R_+-R_-}+\\
&+&\frac{2 i k_- k_0 e^{-i a k_0}}{k_-^2 \sin \left(a k_-\right)+k_0^2 \sin \left(a k_-\right)+2 i k_0 k_- \cos \left(a k_-\right)}\frac{R_+}{R_+-R_-},\\\nonumber
C_8&=&-2 i k_0 e^{-i a k_0}\frac{k_- \left(k_+^2+k_0^2\right) \sin \left(a k_+\right)+2 i k_- k_+ k_0 \cos \left(a k_+\right)+k_+ \left(-\left(k_-^2+k_0^2\right) \sin
   \left(a k_-\right)-2 i k_- k_0 \cos \left(a k_-\right)\right)}{\left(\left(k_-^2+k_0^2\right) \sin \left(a k_-\right)+2 i k_- k_0 \cos
   \left(a k_-\right)\right) \left(\left(k_+^2+k_0^2\right) \sin \left(a k_+\right)+2 i k_+ k_0 \cos \left(a k_+\right)\right)}\times\\
&\times&\frac{R_+R_-}{R_+-R_-}.
\end{eqnarray}
In the limiting case when $n_1\rightarrow1$ and $n_{2,3}\rightarrow0$, we have $\frac{R_+R_-}{R_+-R_-}\rightarrow0$, $\frac{R_-}{R_+-R_-}\rightarrow0$ and $\frac{R_+}{R_+-R_-}\rightarrow1$. From here we conclude immediately that $C_2=C_8=0$. Furthermore, we should go back to the Eq. (\ref{xxx}) that defines the wavefunction to see the behavior of the coefficients inside of barrier. There we see that $(1+jR_+)C_3\sim\mathrm{const}(1+jR_+)\frac{R_-}{R_+-R_-}\rightarrow0$. Similarly we get $(1+jR_+)C_4\rightarrow0$, $(1+jR_-)C_5\sim\mathrm{const}(1+jR_-)\frac{R_+}{R_+-R_-}\rightarrow\mathrm{const}$ and $(1+jR_-)C_6\sim\mathrm{const}(1+jR_-)\frac{R_+}{R_+-R_-}\rightarrow\mathrm{const}$. Therefore, all the quaternionic parts disappear in the limiting case.
We can set the following notation $n_1=\cos\theta$, $n_2=\sin\theta\cos\varphi$ and $n_2=\sin\theta\sin\varphi$. Clearly $\theta=0$ belongs to the pure complex case. We obtained the following Taylor expansion for $\theta\ll1$ and $a,V_0\ll\omega_0$:
\begin{eqnarray}
C_1&=&-i a V_0,~~C_2=a \theta  e^{-i \varphi } V_0,\\
C_3&=&0,~~C_4=0,\\
C_5&=&\frac{V_0}{2 \omega _0}+1,~~C_6=-\frac{V_0}{2 \omega _0}-i a V_0,\\
C_7&=&1-i a V_0,~~C_8=a \theta  e^{-i\varphi } V_0.
\end{eqnarray}
This clearly shows that quaternionic amplitudes necessarily show-up outside of barrier.

\end{document}